\documentclass{ws-procs975x65}

\def\be{\begin{equation}}
\def\ee{\end{equation}}
\def\ea{{\it et al.}\,}

\def\eg{{\it e.g.},\,}

\def\rel{relativistic \,}

\begin{document}

\title{Relativistic Electrons \& Magnetic Fields in Clusters of Galaxies}
%Proceedings\footnote{\uppercase{T}

\author{Yoel REPHAELI\footnote{\uppercase{T}his work is supported by a 
\uppercase{NASA} grant at \uppercase{UCSD}}}

\address{Center for Astrophysics and Space Sciences, \\
University of California, San Diego, \\
La Jolla, CA 92093, USA\\
E-mail: yoelr@mamacass.ucsd.edu}

\maketitle

\abstracts{RXTE and BeppoSAX observations have yielded evidence for the 
presence of a secondary power-law spectral component in the spectra 
of several clusters of galaxies. This emission in clusters with 
extended regions of radio emission is likely to be by \rel electrons 
that are Compton scattered by the CMB. The radio and non-thermal (NT) 
X-ray measurements yield the values of the volume-averaged magnetic 
field and electron energy density in the cluster extragalactic 
environment. These directly deduced quantities provide a tangible basis 
for the study of NT phenomena in clusters.}

\section{Radio \& X-ray Measurements}

Extended radio emission has been detected in some 40 
clusters.\cite{gae}~\cite{ks}
The emission typically extends over a region $\sim 1$ Mpc in size, with 
spectral indices in the range $1.2-1.7$, luminosities $10^{41}-10^{42}$ 
erg/s, and $\sim 1 \,\mu$G {\it equipartition} magnetic fields. From 
Faraday Rotation measurements of radio sources seen through 16 clusters, 
a mean field value of $\sim 5-10\, (\ell/10\ {\rm kpc})^{-1/2}$ $\mu$G, 
where $\ell$ is a characteristic field spatial coherence length, was 
deduced.\cite{ckb}

A direct prediction from the presence of relativistic electrons in 
the IC space is NT X-ray emission from Compton scattering of the 
electrons by the CMB. This emission is expected in the energy range 
of (at least) $\sim 1$ keV -- 10 MeV, with a spectral index that is 
roughly similar to the radio index. Calculations of the predicted 
properties of this emission were carried out long ago.\cite{r77}~\cite{r79} 
Searches for NT IC emission began with archivel analysis of HEAO-1 
data,\cite{r87}~\cite{r88} and continued with CGRO\cite{r94} and 
ASCA\cite{hen} observations, yielding only upper limits on spectral 
power-law components. The improved sensitivity and wide spectral band 
of the RXTE and BeppoSAX allowed a more detailed spectral analysis of 
long exposure measurements that resulted in significant evidence for 
NT emission in five clusters: Coma, A2256, A2319, A119 \& A754 
(references~\refcite{r99}--\refcite{ff03}).    

The level of NT flux deduced from the RXTE measurements is roughly
$\sim 5-8\%$ of the total emission (which is dominated by thermal 
emission). As an example, the RXTE spectrum of the Coma cluster is 
shown in Figure 1; the cluster was observed for $\sim 260$ ks in 1996 
and 2000. Note the spectral overlap between the PCA and the higher 
energy HEXTE experiments, which also have the same FOV, thereby allowing 
simultaneous determination of the thermal parameters, as well as those 
of a possible secondary NT component. (This is a clear advantage over 
BeppoSAX, whose MECS and PDS do not overlap, neither spectrally nor 
spatially.) 

\begin{figure}[t]
\centerline{\psfig{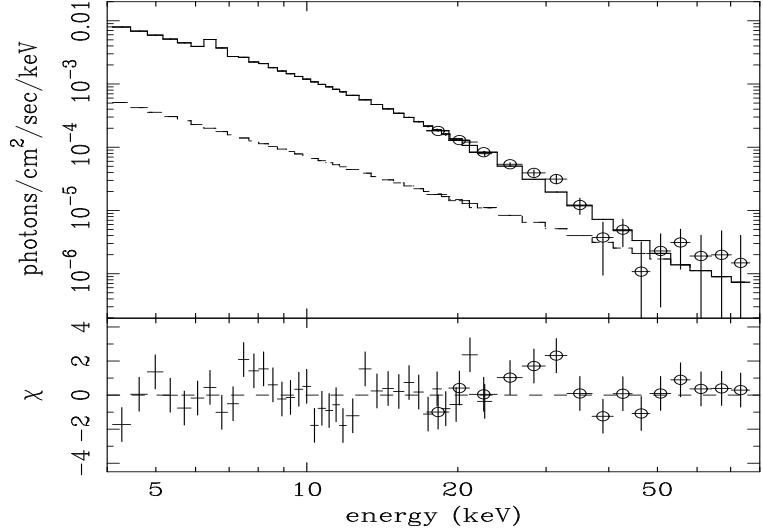}}
\caption{RXTE spectrum of the Coma cluster and folded Raymond-Smith 
($kT \simeq 7.67$), and power-law (index $=2.3$) models (from Rephaeli \& 
Gruber 2002). HEXTE data points are marked with circles and 68\% error 
bars. The total fitted spectrum is shown with a histogram, while the 
lower histogram shows the power-law portion of the best fit. The quality 
of the fit is demonstrated in the lower panel, which displays the observed 
difference normalized to the standard error of the data point.}
\end{figure}
%\eject

\section{Field Strength \& Electron Energy Density}

Radio and NT X-ray measurements yield volume-averaged values of 
the magnetic field in the central $\sim 1$ Mpc region of clusters
$B_{rx} \sim 0.1 - 0.4 \, \mu$G, somewhat lower than values deduced
from Faraday rotation measurements, $B_{fr}$. The mean strength of 
IC fields has direct implications on the feasibility of detecting 
cluster NT X-ray emission, on the electron (Compton-synchrotron) 
loss times, and therefore on the viability of \rel electron models 
(\eg, references \refcite{s99}-\refcite{p01}). Reliable estimates of 
the field are thus quite essential.

Differences between $B_{rx}$ and $B_{fr}$ could, however, be due to 
the fact that the former is a volume-weighted measure of the field, 
whereas the latter is an average along the line of sight, weighted 
by the electron density. In addition, the field and \rel electron 
density would generally be expected to have different spatial 
profiles that could lead to different spatial averages.\cite{gr93} 
More generally, IC field values derived from Faraday rotation 
measurements suffer from various statistical and physical 
uncertainties.\cite{nnr} In particular, the inclusion in the sample 
of {\it cluster} radio sources, whose large contributions to the 
rotation measures could be due to their intrinsic fields, may have 
led to systematic over-estimation of IC fields,\cite{rb} a possibility 
that was recently ruled unlikely.\cite{ens}

When it is assumed (as has been customary in joint radio and X-ray 
analyses) that IC \rel electrons and fields are co-spatial, the deduced 
value of $B_{rx}$ is independent of the volume of the emitting region. 
The electron energy density, $\rho_{e}$, does depend on the source radius 
($R$); integration of the electron energy distribution over energies inferred 
from the observed radio (and X-ray) band yields $\rho_{e} 
\sim 10^{-13 \pm 0.5} (R/1\, Mpc)^{-3}$ erg\,cm$^{-3}$. If 
protons are the main cosmic ray component in clusters (as in galaxies), 
the total particle energy density could be much higher. If so, energetic 
protons can give rise to $\gamma$-ray emission via neutral pion 
decay,\cite{dr} may account for a substantial fraction of the 
\rel electron density through charged pion decays,\cite{bc} and could 
possibly provide appreciable (Coulomb) heating of IC gas within the 
cluster core where the gas (bremsstrahlung) cooling rate can be relatively 
high.\cite{rs}

\end{document}